\documentclass[12pt, preprint]{aastex}
\shorttitle{New insight on IC63/59}
\begin{document}
\title{An Investigation on the Morphological Structure of IC59: \hspace{2 cm} A New Type of Morphology of BRCs?}

\author{Jingqi Miao\altaffilmark{1}, Koji Sugitani\altaffilmark{2}, 
 Glenn, J. White\altaffilmark{3,4}, Richard P. Nelson\altaffilmark{5}}

\altaffiltext{1}{Centre for Astrophysics \& Planetary Science, School of Physical Sciences, University of
Kent, Canterbury, Kent CT2 7NR, UK, J.Miao@kent.ac.uk}

\altaffiltext{2}{Institute of Natural Sciences, Nagoya City University, Mizuho-ku, Nagoya 467-8501, Japan}

\altaffiltext{3}{Centre for Earth, Planetary, Space \& Astronomical Research, The Open University,
Walton Hall, Milton Keynes, MK7 6AA }

\altaffiltext{4}{Space Physics Division, Space Science \& Technology Division,
CCLRC Rutherford Appleton Laboratory, Chilton, Didcot, Oxfordshire, OX11 0QX, UK}

\altaffiltext{5}{School of Mathematical Sciences, Queen Mary College, University of London, Mile End Road,
London E1 4NS, UK}

\begin{abstract}
With references to  recent observational results on the nebula IC59, we applied our 
previously developed Smoothed
Particle Hydrodynamics (SPH) code which is based on Radiative Driven Implosion (RDI) model,  
to investigate the possible formation mechanism for the observed  
morphological structures of  differently shaped BRCs. 
The simulation results confirmed the existence of the 4th type morphology
of BRCs -- type M BRC. We are able to find the necessary condition 
for the appearance of the type M BRC based on the fact that the 
simulated physical properties of the cloud are consistent with observations on IC59. 
More importantly, the prospect of RDI triggered star formation by RDI model 
in all of the observed type M BRCs is ruthlessly eliminated.       
\end{abstract}
\keywords{star: formation -- ISM: evolution -- ISM: HII regions -- ISM: kinematics and dynamics --
 radiative transfer.}

\section{Introduction}
Material at the star-facing surface of dense molecular clouds is intensively ionized 
by the ultraviolet (UV) radiation from nearby OB stars. 
The emission lines from recombination of electrons with ions create a bright rim 
surrounding the star facing surface of the molecular cloud. On the other hand, 
ionization heating drives a strong shock wave into the molecular
cloud so that the compressed gas forms a condensed core behind the bright rim, which is 
the typical structural feature of a  Bright Rimmed Cloud (BRC).   
The Formation of BRCs opens a window for our investigation on the physical processes 
involved in the interaction of UV photons with the gas particles in molecular clouds. 

\begin{figure*}
\rotatebox{-90}{\resizebox{12cm}{16cm}{\includegraphics{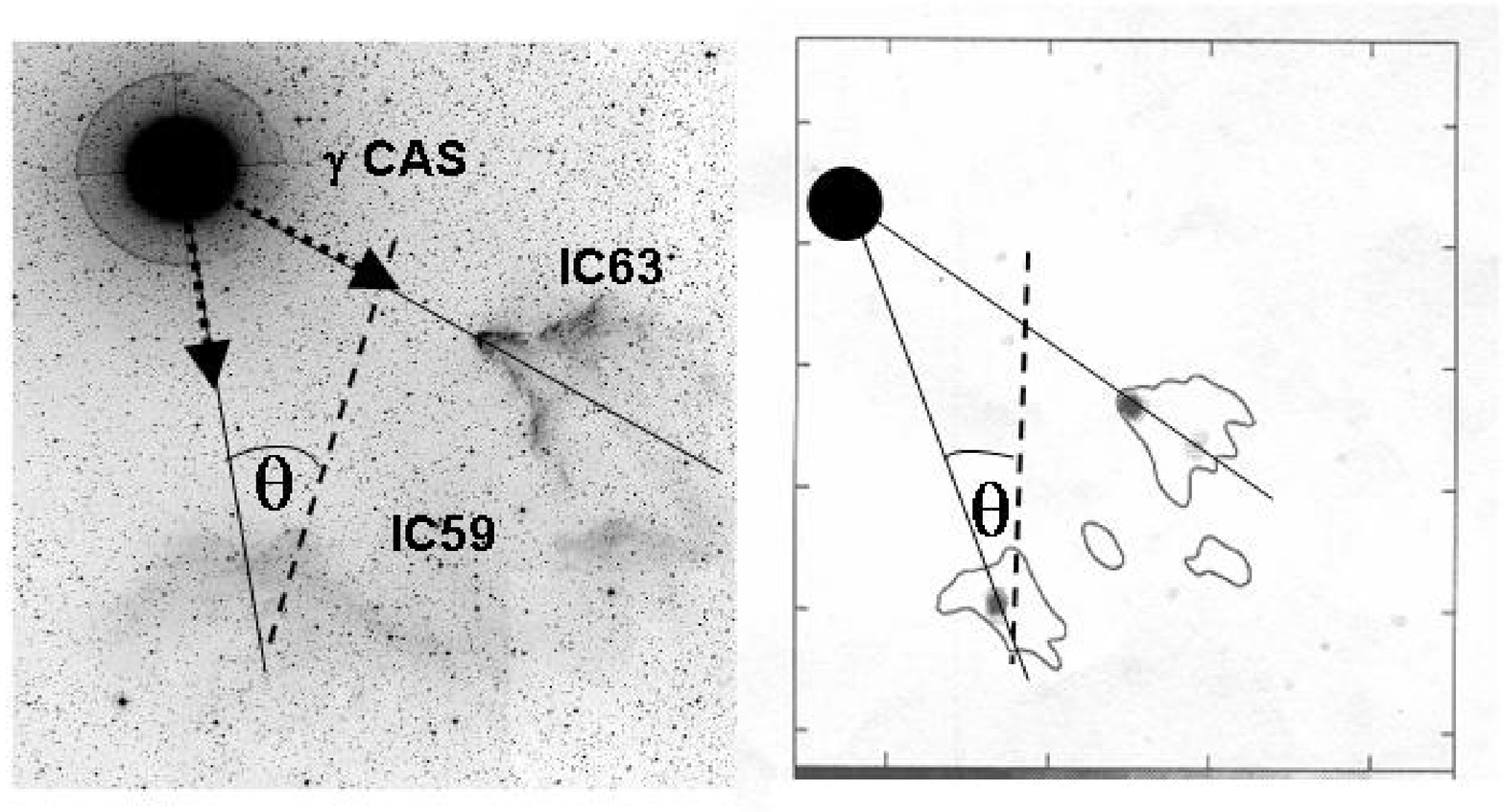}}}
\caption{A geometrical picture of Sh-158 with both IC59 and IC63 surrounds the illuminating 
star $\gamma$ CAS. On the left is theH$_{\alpha}$ images for IC63 and IC59 and one 
right is position of the detected CO emission cores in both clouds \citet{karr}.} 
\label{co}
\end{figure*}

The variety of the morphology of BRCs has interested  
both astronomers and theoreticians since the last decade.
Most of the observed BRCs could be categorized by three different 
morphologies dependent on the curvature of their bright rims: type A, B and C  with an
increased order of degree of the rim curvature \citep{sugitania, sugitanib}. The curved bright 
rims of  type A, B and C BRCs are usually convex  
having the apex of its rim aligned with the center of the condensed core 
 and the radial direction of the illuminating star as shown by the IC63 structure in
both panels of Figure \ref{co}.  
Several theoretical models of the UV radiation effect on the dynamical evolution 
of molecular clouds  have successfully re-produced the formation process
of the above three types of BRCs \citep{lefloch, kessel, williams, miaoa, miaob},
hence we have gained a basic understanding on the mechanism 
of the formation of BRCs with type A, B and C morphologies. 

However, there is another interesting type cometary clouds, whose axial 
line is not aligned with but offset 
from the radial direction of the illuminating star by an angle $\theta$ \citep{osterbrock}, as shown
by the right cap structure of the lower left object in the left panel of Figure \ref{co}.
By a further inspection, one can  find that the observed cometary structure is one of the two 
caps of a bigger cloud structure whose star facing surface is
concave to the radial direction of the illuminating star as shown by the lower
left structure in the left panel of Figure \ref{co}. 
Therefore the symmetrical line  
of any one of the two caps is neither aligned with the radial direction 
of the illuminating star, nor the condensed molecular core behind the 
concave rim as shown by the lower left object in the right panel of Figure \ref{co}.  
This type of cometary structure does not belong to any of the three previously defined 
morphological types. Quite a few known cometary clouds bear the same structural 
feature, e.g.,  the No. 8 source in IC 1396 
\citep{osterbrock} and SFO 80 in RCW 108 \citep{urquhart}. 
The H$_{\alpha}$ images of IC59 and IC63 in the same
region are shown in Figure \ref{co}, which also describes an
offset angle $\theta$ of the symmetrical line of IC59
 from the radial direction of the illuminating star (for the time being we use 
the conventional name IC59 for the right cap structure only and later on we will redefine
the structure range of IC59).     
The physical mechanism for the appearance of this type of morphology  
in molecular clouds, and its relation to the possibility of triggered star formation by 
RDI have not been fully understood.  

Although the offset of the apex of IC59 from the radial direction of the 
illuminating star was tentatively explained as the 
projection effect of a three dimensional space onto a 
two dimensional observation plane \citep{blouin,karr}, 
the observed CO emission core from IC59 by \citet{karr}, 
which usually traces the compressed dense molecular materials
behind the apex of the bright rim of a BRC, 
does not seem to support this explanation. The CO emission core  
presented in the lower part structure in the right panel of  Figure \ref{co} 
reveals a dense core behind the concave rim, which is aligned with the radial 
direction of the illuminating star.
If the offset angle of the apex of IC59 is a consequence of projection effect, the CO emission
core should also be subject to the projection effect so that it 
should appear just behind the apex of the bright rim of the IC59, as the CO emission core 
in IC63 does, as shown in the upper cometary structure in the right panel of Figure \ref{co}.   
Since the center of the CO core in IC59 is aligned with the radial direction of $\gamma$ CAS,
it seems more likely that this type of morphological structure, i.e., 
a condensed molecular core behind the concave rim which has two apexes 
at its two sides, is developed 
during the evolutionary process of a molecular cloud by RDI mode, 
just as the formation of type A, B and C BRCs, whose 
morphology formation mechanism has been well understood through the theoretical
simulations \citep{miaob}. We would categorize this type of structure as 
type M morphology for its geometrical analogy.    
 
In this paper, we will explore the possibility of forming  
 a BRC with a concave rim by RDI mechanism  using 
our previously developed SPH (Smoothed Particle Hydrodynamic) code \citep{miaoa,miaob}, 
and discuss the connection between the appearance of type M BRC
and triggered star formation within it.  
We will structure the paper in the following way: 1.Gather the structural and physical 
properties of IC59 and the illuminating star from available
literature and our own observation; 
2.Present the simulation results which reveals the formation of type M morphology in 
the evolutionary process of a molecular cloud by RDI mode; 
3. Discuss the kinematics behind the formation of type M morphology in BRCs; 
4. Discuss the possibility of RDI triggered star formation in type M BRCs; 
5. Derive a conclusion for result of our investigation.  
\noindent  
\section{Physical properties of IC59 and $\gamma$ CAS}
As seen from Figure \ref{co}, IC59 is one of the pair nebulae in Sh 2-185 and is at a 
distance about 1.3 pc from the illuminating B0 IV star $\gamma$ Cas, which is 
approximately 190 pc away from us. Our understanding on the physical properties of IC59
is gradually improved with the availability of the progressively advanced observing facilities.  
IC59 was once defined as a reflection nebula \citep{osterbrock} for there was little molecular
emission detected \citep{jansena,jansenb} especially when compared with that 
of its neighbor IC63 in the same region. Later the radio continuum and HI study 
by \citep{blouin} revealed the existence of atomic hydrogen in IC59, which were produced through
the dissociation of H$_2$, thus IC59 was taken  
as an emission nebula. The multiwavelength investigation on IC59 \citep{karr}
confirmed the presence of H$_2$ in IC59 and its similarity to IC63 in other 
spectrum features except that an ionization front (a bright rim) in IC59 was not clearly 
detected.

\begin{figure*}
\resizebox{16cm}{14cm}{\includegraphics{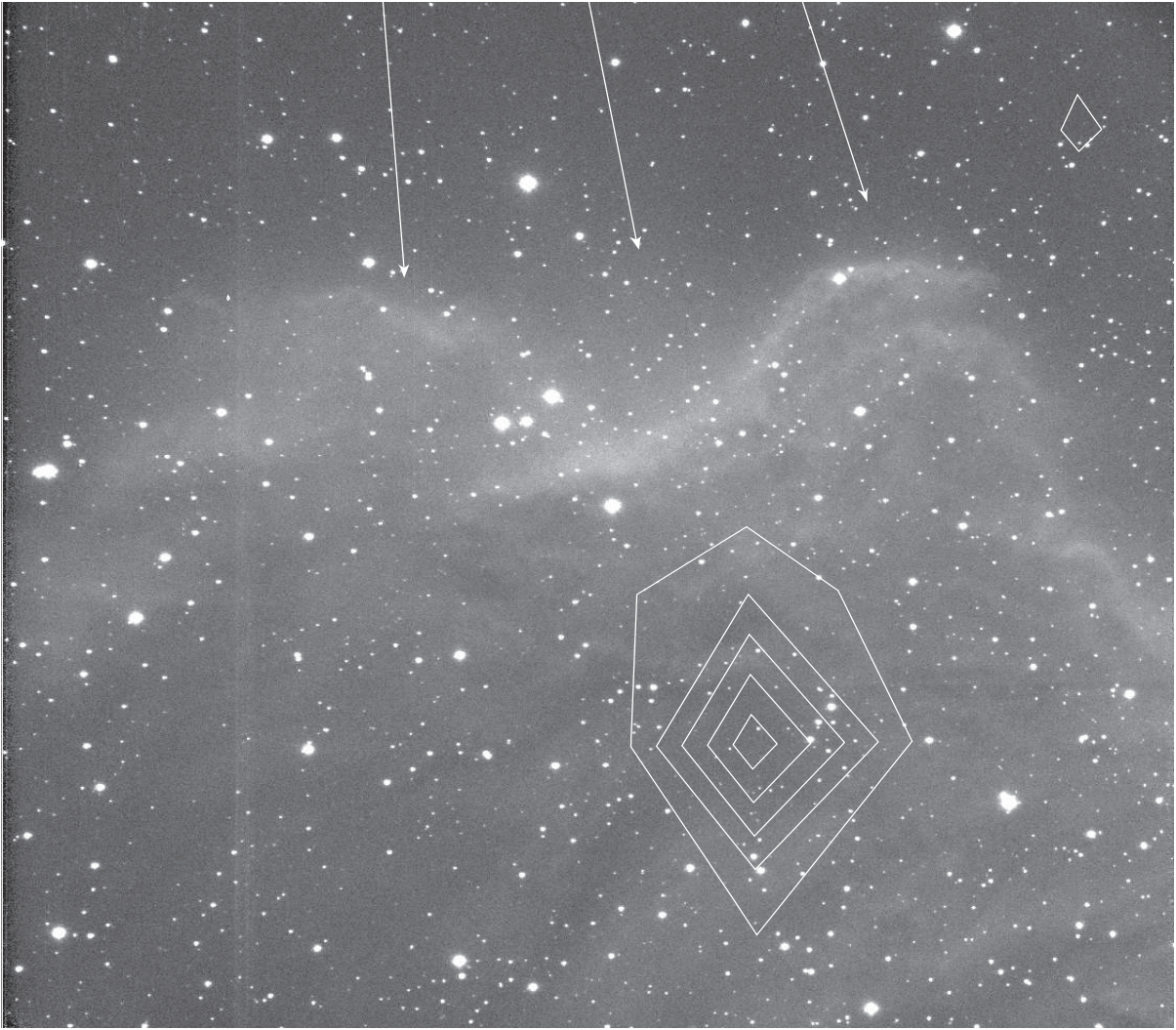}}
\caption{H$\alpha$ image of IC59 overlaid with CO emissions. The area of the image is $\sim$11.5$
\,'$ $\times$ 11.5$\,'$. South is at the top, West to the left.
Arrows indicate the directions of UV light from $\gamma$ Cas.
The pixel scale is 0$\,''$.34 pixel$^{-1}$, providing a field of 
view of 11.5$\,'$ $\times$ 11.5$\,'$.
Three 180 s exposures dithered by 5$\,''$ were taken with H$_{\alpha}$
 narrow band filter.
Dome flat fielding was applied and a combined image was obtained with 
these three images. The CO contours is derived by suming up 5 channels of $V$(LSR) =
 -0.839798, -0.0272369, 0.785339, 1.59790 and 2.41046 km/s. The central LSR velocity of
the integrated intensity map is about 0.8 km/s and the contour interval is 1 K km/s.}
\label{alpha}
\end{figure*}

Recently we obtained H$\alpha$ images  with the Wide Field Grism Spectrograph 
2 (Uehara et al. 1994) and Tektronix 2048 $\times$ 2048 CCD mounted
on the University of Hawaii 2.2-m telescope on 2009 September 10 UT.
 It is seen from Figure \ref{alpha} that around the star facing
side of the nebula, a bright rim
covers the east cap, the concave surface and the west cap. The 
brightest part of the rim is the concave part whose center 
 is aligned with the radial direction of 
the star so that it receives stronger UV radiation than the surfaces
around the two side-caps. Therefore it seems more reasonable to consider the two apexes 
and the concave part as one whole piece of the nebula structure and name it as IC59, 
rather than the east part only. We will use this new definition of IC59 
in our following discussions. With the position of CO emission core in IC59
defined by \citet{karr} and the latest H$_{\alpha}$ image of IC59 we obtained, we 
believe that we have gathered enough evidence to ascertain that the newly
defined  IC59 structure
should be a type of BRC structure with type M morphology, although which has never been 
investigated.

The geometrical and physical parameters we can collect from different literature's  
for the east cap of IC59  are as follows.  
The width of the cap $w \sim $ 0.18 pc; length $l \sim $ 0.12 pc \citep{osterbrock} where the 
dimensions are defined by the convention for BRCs \citep{osterbrock, sugitania, sugitanib}; 
its column density is $3.4\times 10^{17}$cm$^{-2}$ \citep{karr};
the ionization flux at the star facing surface of IC59 is $F_{UV} = 5 \times 10^{9}$cm$^{-2}$s$^{-1}$, 
under the premise that $\gamma$ Cas is a B0 star with a temperature 33,340 K \citep{vacca, karr}.   
An average temperature of the east cap of IC59 is estimated as $T = 590$ K \citep{karr}. 

The above collected properties of IC59 are used as a guidence for the selection of
an appropriate candidate from our simulation experiment in order
for us to address the formation mechanism of 
type M BRC from an initially uniform molecular cloud. We run our SPH code (named as NEWcloud) 
for molecular clouds of different initial conditions. From the results obtained,  
we are able to reveal the possible origin and future evolution of IC59, 
and especially address the possibility of 
RDI triggered star formation within IC59 and other similar  BRCs.        
It is our intention to focus on the formation mechanism of type M BRCs in this 
investigation, therefore
the properties of IC63 is not included in our investigation, 
for which we will address its physical 
properties and evolutionary features in a separate paper in preparation.   

\section{Results and discussions}
\subsection{The formation of type M morphology}
The SPH code we developed \citep{miaoa, miaob} solves the compressible fluid 
dynamic equations, with inclusion of self-gravity of the
cloud, the most important heating and cooling processes plus a chemical network of 12 
basic chemical components in BRCs, which has been fully explained  
in our  previous two papers and will not be repeated here. Interested readers could read 
the above two cited papers to know the details of the code. The radiation fields consists of an 
standard isotropic interstellar background radiation (with photon's 
energy between 6.4 and 13.6 eV)\citep{habing} and a UV radiation field
from nearby star (with photon's 
energy higher than the hydrogen ionizing energy of 13.6 eV). The cloud is initially
situated at the center of a $xyz$ coordinate and the UV radiation is set 
along $-z$ direction and incident on the surface of the upper hemisphere of
the simulated cloud, with a constant energy flux $F_{UV}$ 
specified in the last section.
 
The simulated cloud is initially 
situated in a warm and defuse interstellar medium, which is assumed to compose primarily 
of atomic hydrogen with $n$(HI) = 10 cm$^{-3}$ and $T$ = 100 K \citep{nelson, miaoa}.  
The molecular cloud is of an initial temperature of 60 K, mass of 2 M$_{\sun}$ and 
a radius of 1.4 pc (corresponding to an initial number density 
$n_i$(H$_2$) = 3.48 cm $^{-3}$, which is within the observed range of volume densities of star forming
molecular clouds \citep{heiner}) In all of the simulations presented in the following,
20,000 particles are used.
 
\begin{figure*}
\resizebox{16cm}{14cm}{\includegraphics{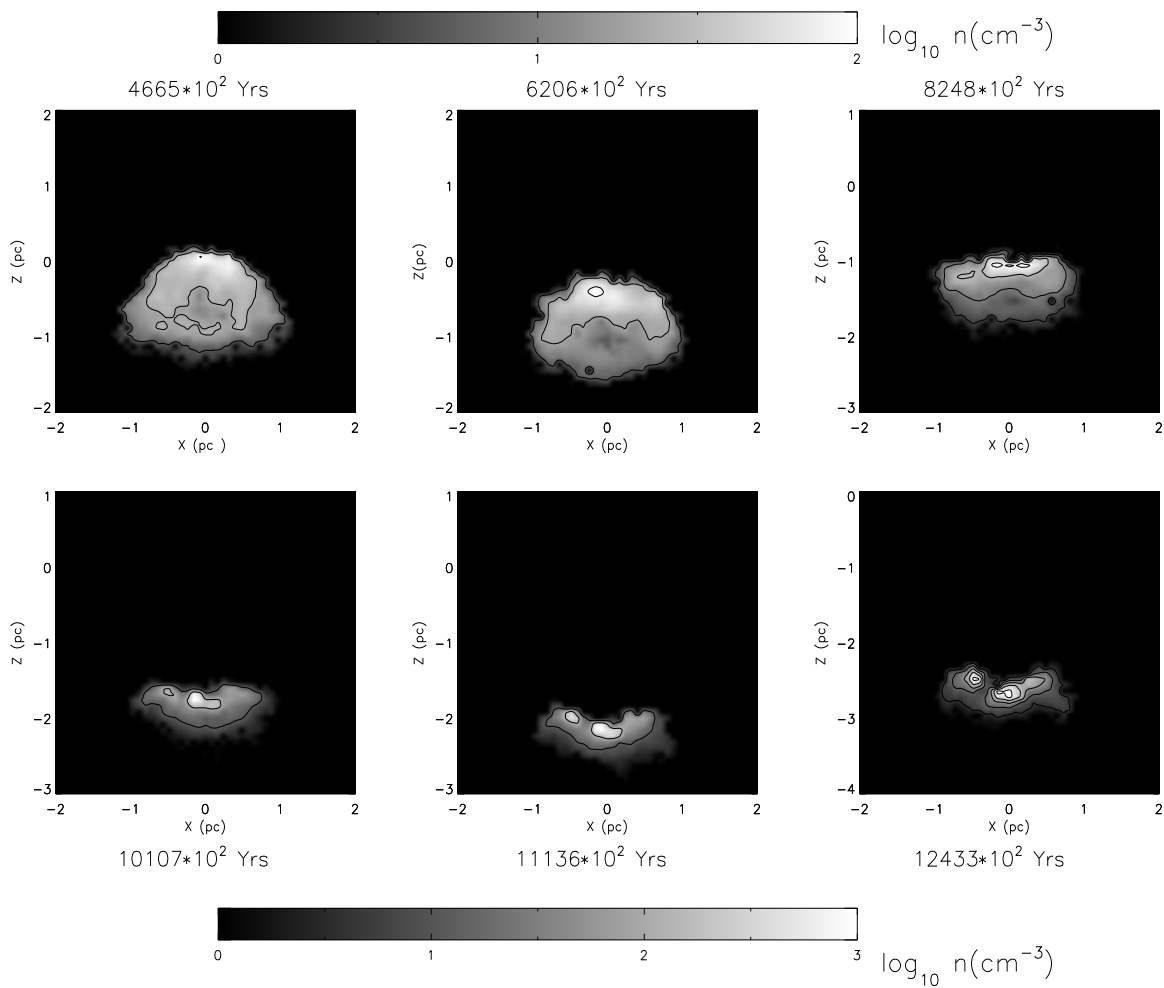}}
\caption{The evolutionary sequence of an uniform molecular cloud with an initial mass of 2 
M$_{\sun}$ and radius of 1.4 pc under the effect of UV radiation field with an ionising photon
flux of $5 \times 10^9$cm$^{-3}$. The UV radiation is
 along the $-z$ direction from the above of the cloud.  
The images and contours are both for number density of
hydrogen atoms within the slice of $\Delta y = 0.056 pc$ centered at $y = 0$. A scale on the numbersity 
is shown on the gray bars.}
\label{den}
\end{figure*}

The hydrogen number density profile in the up-left 
panel in Figure \ref{den} shows that 0.47 My after the radiation fields 
were switched on, the initially spherical cloud evolved into a type A BRC having
 its front surface
(star facing side) squashed from a hemispherical surface and a maximum density 
of 30 cm$^{-3}$ was achieved at the head of the upper hemisphere as a consequence of
UV radiation induced shock propogation into the cloud, whose physical implications have 
been fully understood \citep{bertoldi, lefloch, kessel, williams, miaob}.
The up-middle and -right panels show that the front surface of the cloud gets more
 squashed  at $t =  0.62$ My and then become very flat at $t = 0.82$ My, whilst a 
more and more condensed core appeared with a centre density of 
136 cm$^{-3}$ at t = 0.83 My. 
When $t = 1.01$ My, the front surface of the cloud
started to become concave to the radial direction of the star which 
is above the cloud at positive $z$ axis. A further condensed core formed 
behind the concave surface
and had a centre density of 1127 cm$^{-3}$. The degree of concavity of the front surface
got enhanced at $t = 1.11$ My. It is seen from the
 bottom-middle panel in Figure \ref{den}, that the morphology of 
the cloud structure is very 
similar to the observed morphology of IC59 as shown in both panels in Figure \ref{alpha}, 
 i.e., a concave surface whose center is aligned with $z$ axis (the radial direction of the star), is in between two
caps whose geometrical symmetrical line is offset an angle from the $z$ axis. Thus 
a type M BRC has formed and the centre density of the core behind the 
concave part of the rim is 807 cm$^{-3}$ which is less denser than that at $t = 1.01$ My. 

It is obviously seen that the core behind the concave surface expanded in volume
from $t = 1.01$ to $t = 1.11$ My, therefore the centre density of the core decreased. 
The bottom right panel in Figure \ref{den} shows that the whole 
structure retained a type M morphology but further expanded at $t = 1.24$ My
and the centre density of the core further decreased to 581 cm$^{-3}$. Further
simulation sequences after $t = 1.24$ tells the whole structure contines expansion
and finally expands away  as large piece of very defused cloud  at  $t = 1.7$ My. 
     
Taking any one of two caps in the formed type M BRC at $t = 1.11$ My as an example, 
we can estimate its average column density and then compare it with the observational data.
We take the east cap structure in the middle-bottom panel of 
Figure \ref{den} as the corresponding structure
previously defined as IC59 by astronomers. Simulation result revealed that the 
 structure of the east cap has a mass of 
$3 \times 10^{-4}$ M$_{\sun}$. If we assume the mass 
distributes uniformly in a cap of a sphere of height 0.13 pc
and radius 0.10 pc according to the simulated dimension, which is very close 
to the measurement (0.12, 0.09) pc by 
\citet{osterbrock}, a minimum value of the 
column density $N = 2.2 \times 10^{17}$ cm$^{-2}$ is derived which is consistent with 
the observed value of $3.4 \times 10^{17}$ cm$^{-2}$\citep{karr}. The result for the
west cap is similar to the east cap.  

\subsection{Kinematics for the formation of type M BRCs}     
\begin{figure*}
\resizebox{14cm}{13cm}{\includegraphics{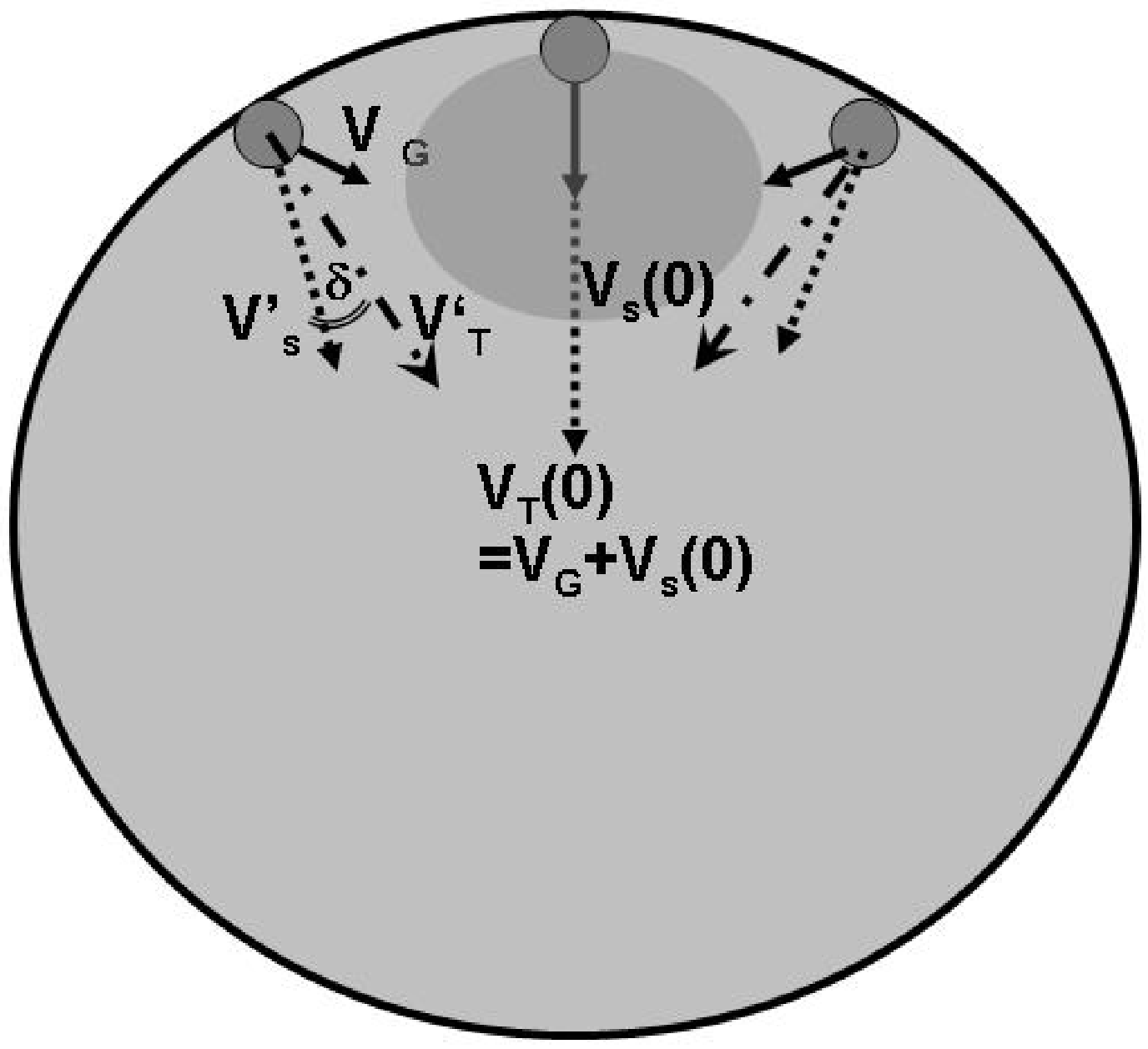}}
\caption{The diagram of velocities of the gas particles at the apex and off-apex point on 
the front surface of the BRC in our simulation.}
\label{vel}
\end{figure*}

A qualitative analysis on the kinematics of the evolution of the cloud would 
reveal a physical picture for the formation of type M BRC. 
The cloud is not strongly bound by self-gravity since it has an initial
Jeans number $\alpha$ (the ratio of the gravitational to the thermal energies) of 
0.025 which is far from the value of 1, the necessary condition for a molecular cloud
to be unstable to gravitational collapse. After the initial stage of the evolution ,
the head of the front hemisphere of the cloud is modestly compressed at $t = 0.62$ My 
so that the gravitational center of the cloud has moved to the head of the upper 
hemisphere as shown in the 
up-middle panel in Figure \ref{den}. The gas particles at the apex and off-apex points 
on the surface of the upper hemisphere have approximately similar gravitational acceleration,
hence similar radial velocity $V_G$ as illustrated in Figure \ref{vel} by the solid arrowed lines.
 One the other hand, 
the UV radiation induced shock velocity $V'_s(\theta)$ (by the dot arrowed lines in 
Figure \ref{vel} ) at the point ($R, \theta$ ) on 
the front surface  is given by \citep{bertoldi,lefloch},
\begin{equation}
V'_s(\theta)= V_s(0)(cos(\theta))^{1/4}
\label{velo-e}
\end{equation}
where $\theta$ is the angle of a surface point from $z$ axis and $ 0
\le \theta \le \pi / 2 $ for all of positions at the front surface 
of the upper hemisphere of the cloud; $V_s(0)$ is the shock velocity 
 at the apex $r=R,\theta = 0 $ and $V_s(0) \sim [F_{UV} / n ]^{1/2}$ with $n$ being the
number density of the site \citep{bertoldi}. 

For the molecular cloud we simulated, because of its
 very low initial density, the shock velocity $V_s(0)$ is one order of magnitude
higher and the radial velocity $V_G$ is much lower than that
in those clouds in which type A, B and C BRCs formed with a high probability
of triggered star formation in their shocked cores \citep{miaob}.  
 Therefore    
 the total velocity $V'_T$ (by the dot-dash arrowed lines in Figure \ref{vel})
of a gas particle at the front surface forms a very 
small angle $\delta$ from the direction of $V'_s$ as shown by Figure \ref{vel},
 so that $V'_T$ is actually 
dominated by the component $V'_S$ which is dependent on the angular distance $\theta$ of
the position from $z$ axis. Consequently $V'_T$ is determined by $\theta$ and will decrease
with $\theta$. Now it is 
clear that the gas particles at the apex of the front surface have the 
highest total velocity $V_T = V'_T(0)$ whose direction is aligned with $-z$ axis as shown
by Figure \ref{vel}.  
After some period of time, the gas particles at or close to the apex of the front surface
are in advance of the particles at two sides of the apex along  $-z$ axis, which results in
formation of a concave surface on the star facing side  
of the molecular cloud as shown by the lower left, middle and right panels in Figure \ref{den}.        

\subsection{Other fingerprints of the simulated cloud}
\begin{figure*}
\resizebox{16cm}{14cm}{\includegraphics{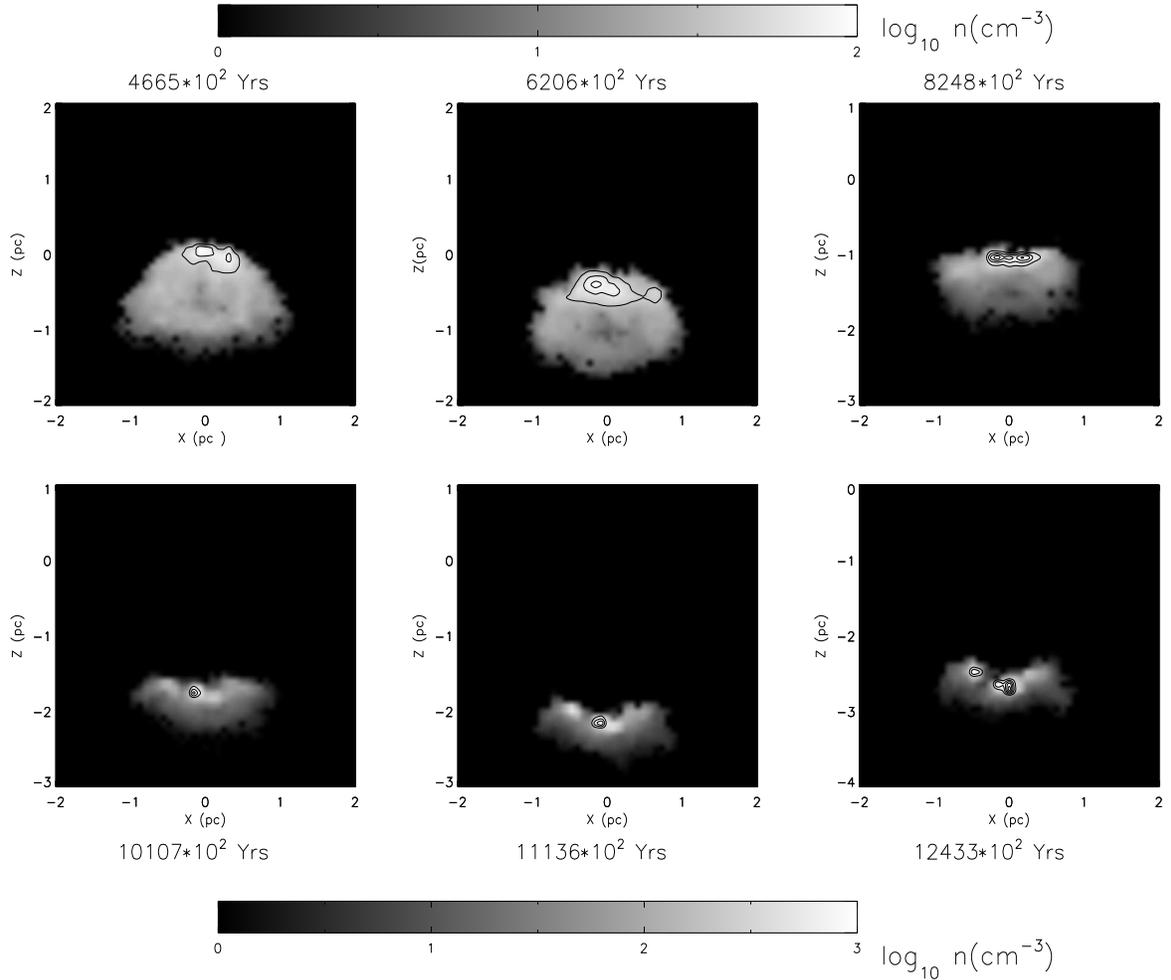}}
\caption{The evolution of the fraction of CO molecules is shown by the contours which is overlaid     
the number density images of the gas particles within the slice of $\Delta y = 0.056 pc$ 
centered at $y = 0$. A scale on the number density images 
is shown on the gray bars.}
\label{denco}
\end{figure*}

Figure \ref{denco} displays a sequence of formation of a CO core  behind the concave
surface. Simulation data reveals that the fractional concentration $X_{CO}$
varied with the centre density of the compressed core since
 CO is a very important ingredient for tracing the dense gas in a molecular
cloud. From $t = 0.47$ to $1.01$ My, the centre concentration of $X_{CO}$ 
 in the compressed core increased from $1.55 \times 10^{-11}$ to 
$7.9 \times 10^{-7}$ and then decreased to 2.2 $\times 10^{-7}$ at 
$t = 1.11$ My, to 3.08 $\times 10^{-8}$ at $t = 1.24$ My. The variation of
the CO fraction  reflects the   
change of the centre density in the cloud structure over the whole evolutionary sequence.
 The position of
the CO core at $t = 1.11$ My is very similar to the observed one as
shown in IC59 structure in lower part of the right panel of Figure \ref{co} \citep{karr}. 

\begin{figure*}
\resizebox{16cm}{14cm}{\includegraphics{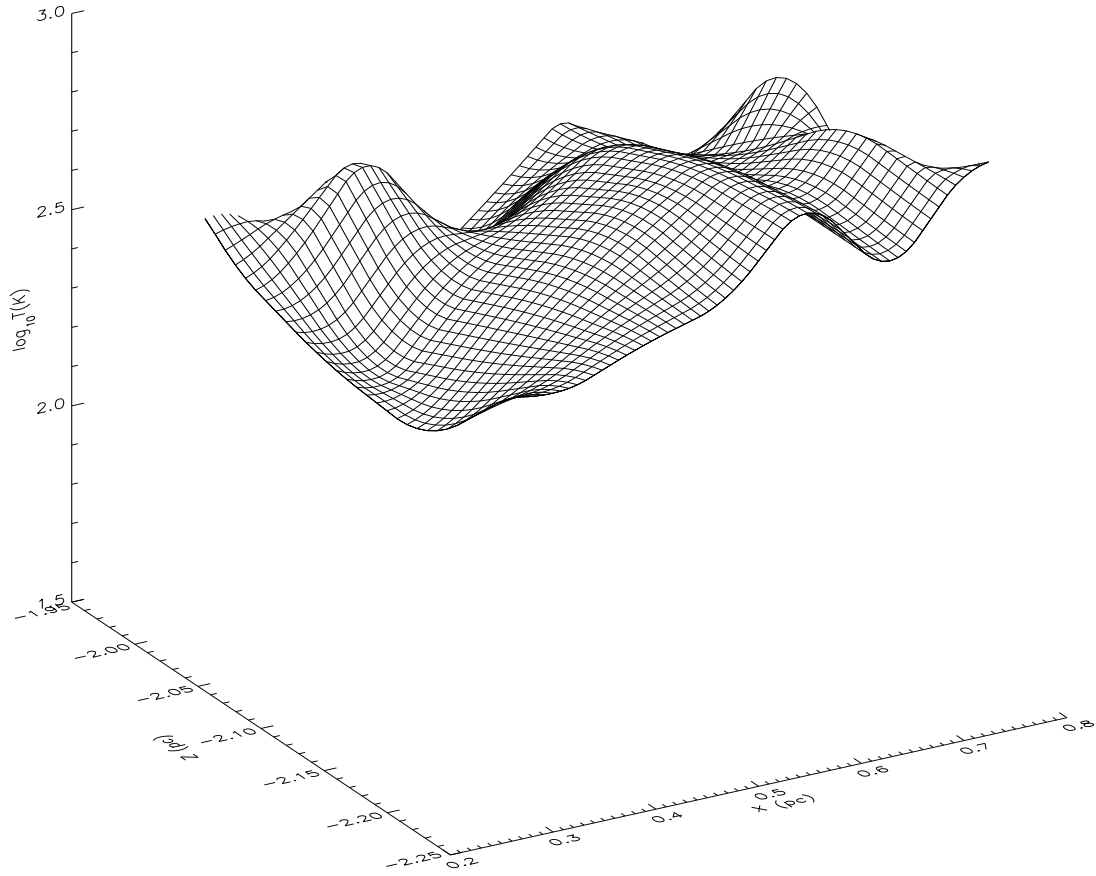}}
\caption{The temperature distribution of the east-cap of IC59 (within the slice of $\Delta y = 0.056 pc$ 
centered at $y = 0$) at $t = 1.11$ My.}
\label{tempcore}
\end{figure*}

In order to investigate the thermal properties of the caps, the east cap was
 chosen for keeping a consistency with the column density
estimation in the last section. Figure \ref{tempcore} presents a 
temperature distribution of the east cap within a slice of $\Delta y = 0.056$ pc centered
at $y = 0$. 
Because of the low initial density of the cloud (3.48 cm$^{-3}$), 
ionizing flux easily penetrates into the cloud structure and
gas particles in the cap are heated dominantly by ionizing hydrogen atoms. 
The average temperature over a region of  $0.43 < x < 0.63$ pc, $-2.08 < z < - 1.95$ pc
is $ ~ 520$ K, which is comparable with the value of 590 K estimated by \citet{karr}, while the
average temperature in the core behind the concave rim is only 65 K due to its higher 
density (807 cm$^{-3}$   ) than that in the cap structure (\~ 40 cm$^{-3}$).   
The dimension of the region with such a temperature distribution is 
also consistent with that was determined by \citet{osterbrock} for the
dimensions of the east cap: a diameter of 0.18 pc and height of 0.12 pc. 
Beyond the front surface of the east cap, the temperature reaches $10^4$, typical to
HII region. Around the front surface of the east cap, recombination of electrons
with hydrogen ions creates the bright rim as shown in Figure \ref{alpha}, although dimmer
than that around the concave part of the rim, where the ionisation/recombination are
most active.    

\subsection{The structure formation and the fate of IC59 }    
Taking the consistent results derived from our simulation 
for the column density, the position of CO core, the dimension 
and mean temperature of the east cap with that from observations, we could
confidentlly construct a whole evolutionary pictrue for the formation of 
the whole structure of IC59: An initially 
uniform and spherical molecular cloud  evolved into a type M BRC after being
exposed to  the ionizing radiation from the nearby star $\gamma$CAS for 1.11 My; the 
observed cap structure is one of its two caps formed at the above specified time, 
where there is no condensed cores inside due to its high temperature state (520 K); the
whole structure will expand away after t = 1.7 My. 
 
centre of the cap  
 Therefore  no IRAS point sources should be observed in both of caps' structures. 
Our simulation results strongly support the comment made by \citet{karr} on the
the nature of the previously reported IRAS point source inside the east cap, i.e., 
'the spuriously detected IRAS point source inside
is merely an unresolved dust feature'. From the simulation result, we conclude that
there is no prospect for triggered star formation in the whole structure of IC59 at all.
We will address a general criteria in the next subsection for the prospect of RDI 
triggered star formation in type M BRCs.         
           
\subsection{Formation of type M BRC and trigger star formation}
In order to find the physical conditions for the formation of type M BRCs and
its relation to the possibility of triggered star formation,
we firstly define the initial position of our simulated molecular cloud in
 Lefloch's two-parameter space, the ionization parameter $\Delta$ and 
recombination parameter $\Gamma$  which are 
expressed in the following forms \citep{lefloch}: 
\begin{eqnarray}
\Delta = \frac{n_i}{n_0} = f_1(R, F_{UV},c_i, n_0)  
\label{parameter1}\\
\Gamma = \frac{\eta \alpha_B n_i R}{c_i} = f_2(R, F_{UV},c_i, n_0)
\label{parameter2}
\end{eqnarray}
where $n_i$ and $n_0$ is the initial densities of ionized and neutral 
hydrogen atoms respectively; $c_i$ is the isothermal sound speed in the ionized gas;    
$\eta \approx 0.2 $ is a parameter to describe the effective thickness of the recombination layer
around the molecular cloud; $\alpha_B$ is the effective recombination coefficient under the
the assumption of 'on the spot' approximation \citep{dyson}; $R$ is the initial radius of the cloud;
$f_1$ and $f_2$ are two functions of the $R, F_{UV}, c_i$ and $n_0$. 
This two dimensional parameter space was divided into five different 
regions depending on the prospects of the evolution of molecular clouds under the effect
of UV radiation. According to lefloch's RDI modeling without inclusion of
the self-gravity of the system, the clouds in region I ($\frac{\Gamma}{\Delta} \le 1.4 \times 10^{-2}$)
and II ($\Delta < 10^{-2} - 10^{-3}$) are too trivial to discuss because the effect of the ionization 
is too weak to produce any noticeable dynamical effects; clouds in region III (defined as 
$\Delta > 2$) is entirely photo-ionized by an R-weak ionization front and there is no
possibility for RDI triggered star formation. 
Recent investigation based on a SPH code developed by the authors of this paper
 has proved that the lower boundary of region III should be shifted to $\Delta > 23$ as
the result of inclusion of the self-gravity of the cloud in the our code \citep{miaob}. 
Classifications of Region IV and
V are not very relevant to the clouds we are interested in this work, therefore will not
be described here. 

The cloud we simulated above (termed as cloud A in the following) has $\Gamma = 21$ 
and $\Delta = 57$, so that its 
initial location in Lefloch's two-parameter space is in region III ($\Delta > 23$), which means that
there should be no prospect for triggered star formation 
in a cloud of an initial mass of 2 M$_\sun$ and a radius of 1.4 pc, while 
F$_{UV} = 5 \times 10^9$ cm$^{-2}$s$^{-1}$. The simulation result presented in the last section
on the future evolution of the IC59 is consistent with the
predicted prospect of the BRCs by the modifiled Lefloch's 2-dimensional
parameter space model. 

In order to show that type M BRC is indeed one of morphologies resulted from UV radiation
effect and type M BRC formation in cloud A is not by chance, we made further
 exploration to a few more molecular clouds of different initial physical conditions.    
We further found another two molecular clouds (termed as B, C clouds in the following)
which would develop type M BRCs and  finally 
expand and split into pieces as cloud A does.  We listed
their initial physical properties in Table \ref{clouds} with the calculated values of 
ionization/recombination parameters.  It is clearly seen that their initial physical status make
them all locate in region III. Following the way we did our exploration,  we can, in principle,
found infinite number of molecular clouds with different physical conditions and under
different strength of UV radiations, which would form a type M BRC by RDI
mode.  Although we are not able to  prove that
all of the molecular clouds located in region III in Lefloch's two-parameter space will form
type M BRCs, we can at least conclude that an initial location in region III in the $\Delta/\Gamma$
 parameter space is a necessary condition for a molecular cloud to develop a type M BRC morphology, i.e, 
a molecular cloud has to be in a very high initial ionization state, as the definition of 
Region III described. Furthermore it is also true that 
the triggered star formation by RDI mode is not possibly to occur 
in all of type M BRCs observed, which may form a good guidance for astronomers who are looking for
the sign of triggered star formation in different BRCs.
   
\section{Conclusion}
Based on our latest observation and 
numeriacl simulation results to the nebula IC59 in Sh-158, we conclude that 
the observed cap structure (so called IC59 in Sh-158) is actually one of the two caps of 
a type M BRC which is a transient morphology from type A morphology to a full expansion, 
which is developed over the evolution of a molecular cloud of an initial mass of 2 M$_\sun$ and
a initial radius of 1.4 pc, under the effect of
UV radiation from nearby star  $\gamma$ CAS. The molecular cloud is at a very high 
initial ionization state in the ionization flash region III in
Lefloch's two dimensional parameter space. Comparison of the simulated physical
properties of the type M BRC with the observations on IC59 structure tells that the observed 
IC59 has been  exposed to the ionizing radiation from  $\gamma$ CAS for about 1.11 My. 
 Simulation result predicts that the observed IC59 structure will continue expanding and
finally split into pieces after another 0.59 My. The future evolution of IC59 revealed by
 numerical result is consistent with the analytical conclusion on the evolutionary 
prospect of a molecular cloud located in region III in the two-parameter space defined by
\citet{lefloch}, i.e., all of the molecular clouds with their initial status in region III
 will be finally evaporated with no prospect for triggered star formation at all. 

Further investigation found more molecular clouds of different initial conditions 
 could evolved into a type M BRC after a type A morphology was developed through
RDI mode. However these molecular clouds share a common character, i.e., no matter how different their
initial masses or radii, their initial conditions make them all locate initially in region III
in Lefloch's two parameter space. Therefore we are able to define a necessary condition 
for a molecular cloud to develop a type M BRC based on the  
 analytical results \citep{lefloch} and our numerical simulations: the molecular
cloud should have the initial physical conditions that make it an initial state in
region III in Lefloch's two-parameter space.      

Although the formation of type M BRC may not account for all of the obseerved offsets of the apexes 
of BRCs to the radial directions of their illuminating stars, 
our simulation results reveal that if there are two
caps connected by a concave front surface in an observed BRC structure and the concave part is
centrally aligned with the radial direction of the star, then the offset of the apex of the 
observed caps' structure to the radial direction of the star may be
caused by the formation of a type M BRC. In a seperated paper, we will address some other reasons which
could possibly cause the offset of the apex of a single structure BRC to the radial direction of the illuminating
star. More importantly, the presented simulation results in this paper
reveal that triggered star formation will not occur in the observed type M BRC,
neither in its cap structures nor behind the concave front 
surface. 

The above results based on our simulations strongly support the argument made by
\citet{karr} on the nature of the structure of IC59: the previously detected potential example 
of triggered star formation is spurious and the observed IRAS source(s) in IC59 is merely 
an unresolved dust feature.

\clearpage

\clearpage

\clearpage

\clearpage

\clearpage

\clearpage

\clearpage
\begin{deluxetable}{lllllllll}
\tablecolumns{9}
\tablewidth{0pc}
\tablecaption{Parameters of clouds which form type M BRC}
\tablehead{
\colhead{Cloud} & \colhead{$\Delta$} &  \colhead{$\Gamma$} & $n_0(cm^{-3})$ & 
\colhead{R(pc)} & \colhead{Mass(M$_\sun$)} & \colhead{Region}} 
\startdata
A & 57 & 21 & 3.48 & 1.4  & 2 & III \\ 
B & 52 & 18 & 4.37 & 1.03  & 1 & III \\
C & 35  & 13 & 8.8 & 0.56  & 0.4 & III \\
\enddata
\label{clouds}
\end{deluxetable} 

\end{document}